\documentclass[
]{ceurart}

\usepackage{listings}
\usepackage{xurl}
\usepackage{nowidow}

\counterwithout{figure}{section}

\begin{document}

\copyrightyear{2024}
\copyrightclause{Copyright for this paper by its authors.
  Use permitted under Creative Commons License Attribution 4.0
  International (CC BY 4.0).}

\conference{Edge AI meets Swarm Intelligence Technical Workshop,
  September 18, 2024, Dubrovnik, Croatia}

\title{Not All RDF is Created Equal: Investigating RDF Load Times on Resource-Constrained Devices}

\author[1]{Piotr Sowiński}[%
orcid=0000-0002-2543-9461,
email=piotr.sowinski.dokt@pw.edu.pl,
]
\cormark[1]
\address[1]{Warsaw University of Technology, Pl. Politechniki 1, 00-661 Warsaw, Poland}

\author[2]{Anh Le-Tuan}[%
orcid=0000-0003-2458-607X,
email=anh.letuan@tu-berlin.de,
]
\address[2]{Open Distributed Systems, Technical University of Berlin, 10587 Berlin, Germany}

\author[3]{Paweł Szmeja}[%
orcid=0000-0003-0869-3836,
email=pawel.szmeja@ibspan.waw.pl,
]
\address[3]{Systems Research Institute, Polish Academy of Sciences, ul. Newelska 6, 01-447 Warsaw, Poland}

\author[1]{Maria Ganzha}[%
orcid=0000-0001-7714-4844,
email=maria.ganzha@pw.edu.pl,
]

\cortext[1]{Corresponding author.}

\begin{abstract}
  As the role of knowledge-based systems in IoT keeps growing, ensuring resource efficiency of RDF stores becomes critical. However, up until now benchmarks of RDF stores were most often conducted with only one dataset, and the differences between the datasets were not explored in detail. In this paper, our objective is to close this research gap by experimentally evaluating the load times of eight diverse RDF datasets from the RiverBench benchmark suite. In the experiments, we use five different RDF store implementations and several resource-constrained hardware platforms. To analyze the results, we introduce the notion of relative loading speed (RLS), allowing us to observe that the loading speed can differ between datasets by as much as a factor of 9.01. This serves as clear evidence that ``not all RDF is created equal'' and stresses the importance of using multiple benchmark datasets in evaluations. We outline the possible reasons for this drastic difference, which should be further investigated in future work. To this end, we published the data, code, and the results of our experiments.
\end{abstract}


\maketitle

\vspace{-1cm}
\section{Introduction}
\vspace{-4mm}

Improving the resource efficiency of knowledge-based systems is a crucial research question motivated by several real-life use cases.
Examples include scaling up to meet the challenges of Big Data in the Cloud~\cite{lehmann2017distributed,le2013elastic}, or enabling swarm intelligence in small IoT or Edge devices~\cite{le2020pushing}. 
This issue is especially important in the second setting, where available resources are always limited for those devices due to their small size and low power requirements.

At the same time, enriching IoT systems with knowledge is the leading way to tackle the issue of IoT interoperability. Both the already established and more recent interoperability approaches rely on the Resource Description Framework (RDF) to model knowledge, for example: the Web of Things W3C family of standards~\cite{w3cwot}, the FIWARE platform~\cite{cirillo2019standard}, and the many supporting ontologies and tools~\cite{ganzha2017semantic,nagasundaram2022semantic}. Such IoT use cases demand maximum efficiency from the used RDF software. However, assessing which RDF tools are the most efficient is not a trivial task.

Several benchmarks were proposed for evaluating RDF stores~\cite{pan2018survey} and many such studies were conducted over the years. 
However, as shown in the literature review below, these investigations most often use only one dataset, or focus on a single, narrow use case. 
Meanwhile, IoT systems (and, more generally, Cloud-Edge-IoT systems) spread to more and more application areas, and the tasks given to IoT devices are increasingly varied. Therefore, it is not reasonable to evaluate the performance of RDF processing software only on a single dataset, as it would not be representative of the multitude of potential applications of next-generation systems. Furthermore, to inform future software design choices, the differences between real-life RDF datasets must be better understood and linked to their influences on performance results. One such performance metric is the speed of loading RDF data into an RDF store. For most query engines, the data must be loaded and indexed before any queries are executed, making loading an important step in many semantic data processing pipelines, especially if the data changes quickly. Here, the speed with which the RDF store can ingest new data can become a bottleneck.

In this contribution, we explore the outlined research gap in the context of loading data into RDF stores. That is, we experimentally evaluate the RDF loading performance of five RDF stores with eight diverse datasets from the RiverBench benchmark suite~\cite{sowinski2023riverbench}, running on a selection of modern single-board computers. In line with the observed research gap, particular focus is placed on analyzing the differences in performance between the datasets.

This paper is organized as follows. Section~\ref{sec:background} presents the background of the study and the related works. Section~\ref{sec:methodology} describes the methodology of the conducted experiments, while Section~\ref{sec:results} presents the experimental results. Section~\ref{sec:discussion} discusses the results and outlines future work directions. Finally, Section~\ref{sec:conclusion} concludes the paper.

\vspace{-5mm}
\section{Background and Related Work}
\label{sec:background}
\vspace{-3mm}

To provide context for this study, in this section we present an overview of selected representative studies that reported RDF store load times. The overview focuses in particular on the datasets used in the benchmarks and how their impact on load times was discussed.

Many RDF store benchmarks consist of two main parts: evaluating the time taken to load an RDF dataset into the store, and the time needed to run a set of predefined queries. Often, such evaluations are limited to a single benchmark dataset~\cite{neumann2010x,cudre2013nosql,lampo2010benchmark,nikolic2015rdf,punnoose2012rya,cheng2015scale,demartini2011bowlognabench}. Especially popular in this context are the Berlin SPARQL Benchmark (BSBM)~\cite{bizer2009berlin} and the Lehigh University Benchmark (LUBM)~\cite{guo2005lubm}. Both BSBM and LUBM rely on synthetic datasets of arbitrary length, generated to a predefined specification. Although the generators are capable of outputting datasets of varying complexity (i.e., different distributions of relations), they only produce datasets pertaining to one specific use case, with one ontology behind it. The strength of these benchmarks lies in their query diversity, which reliably tests different RDF store capabilities. However, in terms of dataset variety, they are relatively lacking.

It must be noted that using a single dataset for the evaluation makes sense if we are interested in a system's performance only in the context of a specific use case. 
For example, the extensive Europeana RDF store report~\cite{haslhofer2011europeana} evaluates RDF stores specifically to determine which would best suit the Europeana project. Such evaluations were also performed, for example, in the smart city domain~\cite{bellini2018performance} and for IoT applications~\cite{le2020pushing}. These investigations are informative within the context of their application area, but from this we cannot assess how these systems would perform in other use cases.

We have also identified three use case-specific studies which included more than one dataset. Jakobsen et al.~\cite{jakobsen2015optimizing} compared three different RDF representations for data cubes to assess how efficiently they can be processed by RDF stores. However, in terms of loading time, they only observe that larger datasets (measured in triples) take longer to load. Therefore, the impact of other characteristics of datasets on the loading time is not considered. In BioBenchmark Toyama 2012~\cite{wu2014biobenchmark} the authors used five biomedical datasets with varying sizes and representations. They noted that the loading time is mainly influenced by the dataset's size and its representation format (number of files into which it was divided). They hypothesized that differences in the structure of the data may also have some influence, but this was not explored further. Finally, the Geographica 2 benchmark report~\cite{ioannidis2021evaluating} included three different geospatial datasets, and discussed the differences between them at length. For instance, the authors found that the number of predicates in a dataset or the size and content of literals both have a large impact on load times, highlighting the advantages and disadvantages of specific systems.

Evaluations that include several datasets from multiple application areas are relatively rare. One such example is the paper on RDF-3X~\cite{neumann2010rdf}, which used three different real-life datasets. Here, the different datasets serve essentially as test cases. Although they do provide a more reliable view of how RDF-3X would perform in various real-life tasks, the differences between the datasets themselves are not discussed. The same approach (also with three datasets) is used in the RDFBroker study~\cite{sintek2006rdfbroker} and by Voigt et al.~\cite{voigt2012yet} (with four datasets). Finally, Ben Mahria et al.~\cite{ben2021empirical} used 17 datasets in their evaluation, making it very diverse. They then ranked different RDF stores based on the number of datasets in which they perform best. However, their discussion only focuses on the impact of dataset size (in triples) on loading time.

The need for diverse datasets in RDF store evaluations was previously acknowledged by Pan et al. in their survey of RDF benchmark datasets~\cite{pan2018survey}, which notes that ``Different RDF datasets and queries often require different storage solutions. Existing RDF storage methods have advantages and disadvantages in different scenarios''. A similar observation was made by Lampo et al.~\cite{lampo2010benchmark}, who underline the importance of having datasets with different sizes and correlations between nodes. However, as demonstrated above, this is rarely applied in practice. We could find only one study (the Geographica 2 report~\cite{ioannidis2021evaluating}) which discussed the structural differences between datasets and their impact on performance, albeit only within the domain of geospatial data. 
Clearly, this is an under-explored topic, warranting further investigation.

Our recent contribution to at least partially solving this problem is the RiverBench benchmark suite~\cite{sowinski2023riverbench}. It offers a collection of well-described and diverse datasets for RDF system benchmarking. The \href{https://w3id.org/riverbench/v/2.0.1/}{2.0.1 release of RiverBench} also includes benchmark task definitions that specify how to run benchmarks and measure their results. Among the tasks defined in RiverBench is \href{https://w3id.org/riverbench/v/2.0.1/tasks/flat-rdf-store-loading}{\texttt{flat-rdf-store-loading}}, which concerns loading data into RDF stores.

\vspace{-5mm}
\section{Methodology}
\label{sec:methodology}
\vspace{-4mm}

This section outlines the methodology used for the conducted experiments, including the tested RDF stores, hardware platforms, datasets, benchmarking procedures, and metrics.

\vspace{-3mm}
\subsection{RDF Stores}
\vspace{-3mm}

Table~\ref{tab:stores} lists the RDF store implementations used in the experiments, which include JenaTDB2, RDF4J Native Store, RDF4J LMDB, Virtuoso, and RDF4Led. Jena TDB2 is a component of Apache Jena that provides efficient, scalable, and transactional storage and retrieval of RDF data for Semantic Web and Linked Data applications. RDF4J is a Java framework for RDF data management that offers two main storage solutions: RDF4J Native Store, which uses native disk-based storage, and RDF4J LMDB, which integrates with the high-performance Lightning Memory-Mapped Database (LMDB). Virtuoso~\cite{erling2009rdf} is a high-performance, multi-model database management system that supports RDF and SPARQL for Semantic Web and Linked Data applications, implemented in C. Finally, RDF4Led~\cite{le2020pushing} is a lightweight, high-performance RDF storage and processing engine designed specifically for embedded devices and environments with constrained resources. For RDF4J LMDB, native bindings for LMDB were used from the LWJGL library version 3.3.3. For Virtuoso, the official Docker image was used. To test the Java-based RDF stores, Eclipse Temurin version 21.0.3+9 was used as the Java Runtime Environment.

\vspace{-3mm}
\begin{table}[htb]
  \caption{RDF store implementations used in the experiments.}
  \label{tab:stores}
  \begin{tabular}{lcccc}
    \toprule
    \textbf{RDF store} & \textbf{Impl. language} & \textbf{Store type} & \textbf{Version} \\
    \midrule
    Apache Jena TDB2 & Java & Native & 5.0.0 \\
    RDF4J Native Store & Java & Native & 4.3.11 \\
    RDF4J LMDB & Java, C & Memory-mapped key-value & 4.3.11 \\
    Virtuoso Open Source & C & Column-wise & 7.2.12 \\
    RDF4Led & Java & Native & -- \\
    \bottomrule
  \end{tabular}
\end{table}

\vspace{-3mm}
For all RDF stores, the default index settings were used. For Apache Jena TDB2 the default triple indexes are: SPO, POS, OSP. In RDF4J there is no explicit default for the indexing strategy, and thus we used the one suggested in the documentation, with three indexes: SPOG, OSPG, PSOG. For Virtuoso Open Source 7.x, two full and three partial column-wise indexes are used: PSOG (primary key), POGS (bitmap index), SP, OP, GS. Finally, RDF4Led uses SPO, POS, and OSP indexes.

\vspace{-4mm}
\subsection{Hardware Platforms}
\vspace{-3mm}

Table~\ref{tab:hardware} lists the hardware platforms used in the experiments. The study only includes devices with 64-bit CPUs, as recent versions of Virtuoso do not support 32-bit platforms. Nevertheless, modern single-board computers typically use 64-bit CPUs -- an example of such is the Raspberry Pi, starting from version 3. For all devices a 128\,GB MicroSD card with an ext4 filesystem was used as storage. The default kernel and OS configuration settings were used. The only exception was the Raspberry Pi 4B, for which kernel parameters \texttt{cgroup\_enable=memory swapaccount=1} were added to allow for limiting the memory available to Docker containers, as required by the benchmark setup (Section~\ref{sec:procedure}).

\vspace{-5mm}
\begin{table}[htb]
  \caption{Hardware platforms used in the experiments.}
  \label{tab:hardware}
  \begin{tabular}{lcccc}
    \toprule
    \textbf{Device} & \textbf{CPU} & \textbf{RAM} & \textbf{OS} & \textbf{Linux kernel} \\
    \midrule
    Raspberry Pi 3B & 4x Cortex-A53, 1.2\,GHz & 1\,GB & Raspbian 12 & 6.6.20 \\
    Raspberry Pi 4B & 4x Cortex-A72, 1.8\,GHz & 8\,GB & Ubuntu 20.04.6 & 5.4.0 \\
    Raspberry Pi 5 & 4x Cortex-A76, 2.4\,GHz & 8\,GB & Ubuntu 23.10 & 6.5.0 \\
    \bottomrule
  \end{tabular}
\end{table}
\vspace{-5mm}

\subsection{Benchmark Datasets}
\vspace{-3mm}

The latest version of RiverBench (2.0.1) contains 12 datasets in total. RiverBench also offers specific \emph{profiles} that are essentially subsets of these datasets, meeting certain technical criteria (e.g., datasets using only triples, or datasets strictly conformant with standard RDF 1.1). To ensure a fair comparison, a benchmark profile had to be selected such that all tested triple stores supported all datasets in the profile. Firstly, datasets using named RDF graphs (or RDF quads) could not be used, as they are not supported by RDF4Led. Secondly, datasets using RDF-star were also excluded, as this unofficial feature is not supported by both Virtuoso and RDF4Led. These selection criteria are satisfied by RiverBench's \href{https://w3id.org/riverbench/v/2.0.1/profiles/flat-triples}{\texttt{flat-triples}} profile, which includes 8 datasets. Table~\ref{tab:datasets} presents an overview of the datasets, along with their unique subject, predicate, and object counts. Additionally, for each dataset we calculated the mean number of bytes per triple (bpt), when encoded in the N-Triples format. This metric serves to highlight the datasets' diversity. Further metrics about each dataset are available on the datasets' documentation pages in RiverBench (links can be found in Table~\ref{tab:datasets}).

\vspace{-5mm}
\begin{table}[htb]
  \caption{RiverBench datasets used in the experiments. The datasets were taken from the \href{https://w3id.org/riverbench/v/2.0.1/profiles/flat-triples}{\texttt{flat-triples} (version 2.0.1)} benchmark profile. Mean bpt is the mean number of bytes per triple in the dataset, when encoded in the N-Triples format.}
  \label{tab:datasets}
  \begin{tabular}{lrrrrr}
    \toprule
    \textbf{Dataset} & \textbf{Triples} & \textbf{Subjects} & \textbf{Predicates} & \textbf{Objects} & \textbf{Mean bpt} \\
    \midrule
    \href{https://w3id.org/riverbench/datasets/assist-iot-weather/1.0.2}{assist-iot-weather} & 80,646,970 & 12,623,016 & 12 & 7,015,794 & 189.79 \\
    \href{https://w3id.org/riverbench/datasets/citypulse-traffic/1.0.2}{citypulse-traffic} & 157,773,564 & 43,826,439 & 9 & 43,839,219 & 232.20 \\
    \href{https://w3id.org/riverbench/datasets/dbpedia-live/1.0.2}{dbpedia-live} & 21,831,109 & 9,520,399 & 14,752 & 5,922,513 & 157.41 \\
    \href{https://w3id.org/riverbench/datasets/digital-agenda-indicators/1.0.2}{digital-agenda-indicators} & 11,669,016 & 1,440,414 & 10 & 754,963 & 247.55 \\
    \href{https://w3id.org/riverbench/datasets/linked-spending/1.0.2}{linked-spending} & 55,097,866 & 2,475,928 & 708 & 8,085,639 & 214.00 \\
    \href{https://w3id.org/riverbench/datasets/lod-katrina/1.0.2}{lod-katrina} & 179,128,407 & 38,479,105 & 10 & 38,503,088 & 208.09 \\
    \href{https://w3id.org/riverbench/datasets/muziekweb/1.0.2}{muziekweb} & 36,195,263 & 2,450,357 & 46 & 7,983,490 & 134.73 \\
    \href{https://w3id.org/riverbench/datasets/politiquices/1.0.2}{politiquices} & 159,957 & 35,546 & 9 & 54,763 & 142.09 \\
    \bottomrule
  \end{tabular}
\end{table}
\vspace{-4mm}
Each dataset was downloaded in the full flat distribution (as a single long N-Triples file) from the RiverBench website. To allow measuring the loading speed as it changes over time, the datasets were split in 50,000-triple batches, discarding the last batch if it was shorter than 50,000 triples. Table~\ref{tab:datasets} presents the statistics of datasets before this trimming was performed. The batches were saved as N-Triples files and used as input for the evaluated RDF stores. This prepared data is available on Zenodo~\cite{benchmark_data}.

\vspace{-3mm}
\subsection{Benchmark Procedure and Metrics}
\label{sec:procedure}
\vspace{-3mm}

The benchmark was executed in a Docker container which contained both the system under test (the RDF store) and the benchmark driver. The driver was implemented in Java, and is a significantly modified version of the code used in the evaluation of RDF4Led~\cite{le2020pushing}. For RDF stores implemented in Java, their Java APIs were used to construct the store and request loading data from disk. For Virtuoso, the store was instantiated as a background Linux process, and load requests were sent to it via console scripts and the \texttt{isql} command. The code used for the benchmark is available on \href{https://github.com/Ostrzyciel/rdf4led-riverbench}{GitHub} and Zenodo~\cite{code}.

The benchmark container was run with a limited amount of system memory to ensure a consistent testing environment. For the Pi 3B, 500\,MB were allocated for the container, while for the Pi 4B and the Pi 5 two variants were tested: 1\,GB and 2\,GB of RAM. The benchmark driver then set up the store and began loading the batches of triples into the store, always in the same order (as a flat RDF triple stream, according to the RDF Stream Taxonomy~\cite{sowinski2024rdf}). After each batch was completed, the metrics were recorded. The most important collected metric is the loading time, which is simply the time taken to load a batch of 50,000 triples. In the results analysis below, we use the loading speed (LS) metric, which is the loading time divided by the number of triples in the batch. Additionally, performance metrics were collected, such as RAM usage, CPU time usage, and RDF store size on disk. A benchmark run was aborted if either (1) the RDF store crashed -- for example due to an out-of-memory error, or (2) the loading speed for the most recent batch dropped below 80 triples per second. This threshold was chosen to be the same as in the previous evaluation of RDF4Led~\cite{le2020pushing}.
This benchmark procedure was contributed to RiverBench's \href{https://w3id.org/riverbench/v/2.0.1/tasks/flat-rdf-store-loading}{\texttt{flat-rdf-store-loading}} task, version 2.0.1.

\vspace{-5mm}
\section{Experimental Results}
\label{sec:results}
\vspace{-3mm}

In what follows, we highlight the most important results obtained from the experiments. The full results are openly available on Zenodo under the CC BY 4.0 license~\cite{benchmark_results}. The published data also includes other performance metrics (RAM usage, CPU time, store size on disk) that are not discussed here.

Additionally, this benchmark was reported semantically as a \href{https://w3id.org/np/RAyFZlqsYQ_w-j5cah_gI8WBIZxiVSM4ocWHD_tnyjLxs}{nanopublication using the RiverBench vocabulary}. Nanopublications are small units of scientific knowledge published in RDF~\cite{kuhn2013broadening}, such as measurement results or opinions. This nanopublication includes the information on the used RiverBench profile and task, the tested triple stores, and the measured metrics. This mechanism is used by RiverBench to gather benchmark results from the community and aggregate them on the benchmark website.

\vspace{-4mm}
\subsection{Comparison of RDF Stores and Platforms}
\vspace{-2mm}

\begin{figure}[ht!]
  \centering
  \includegraphics[width=\linewidth]{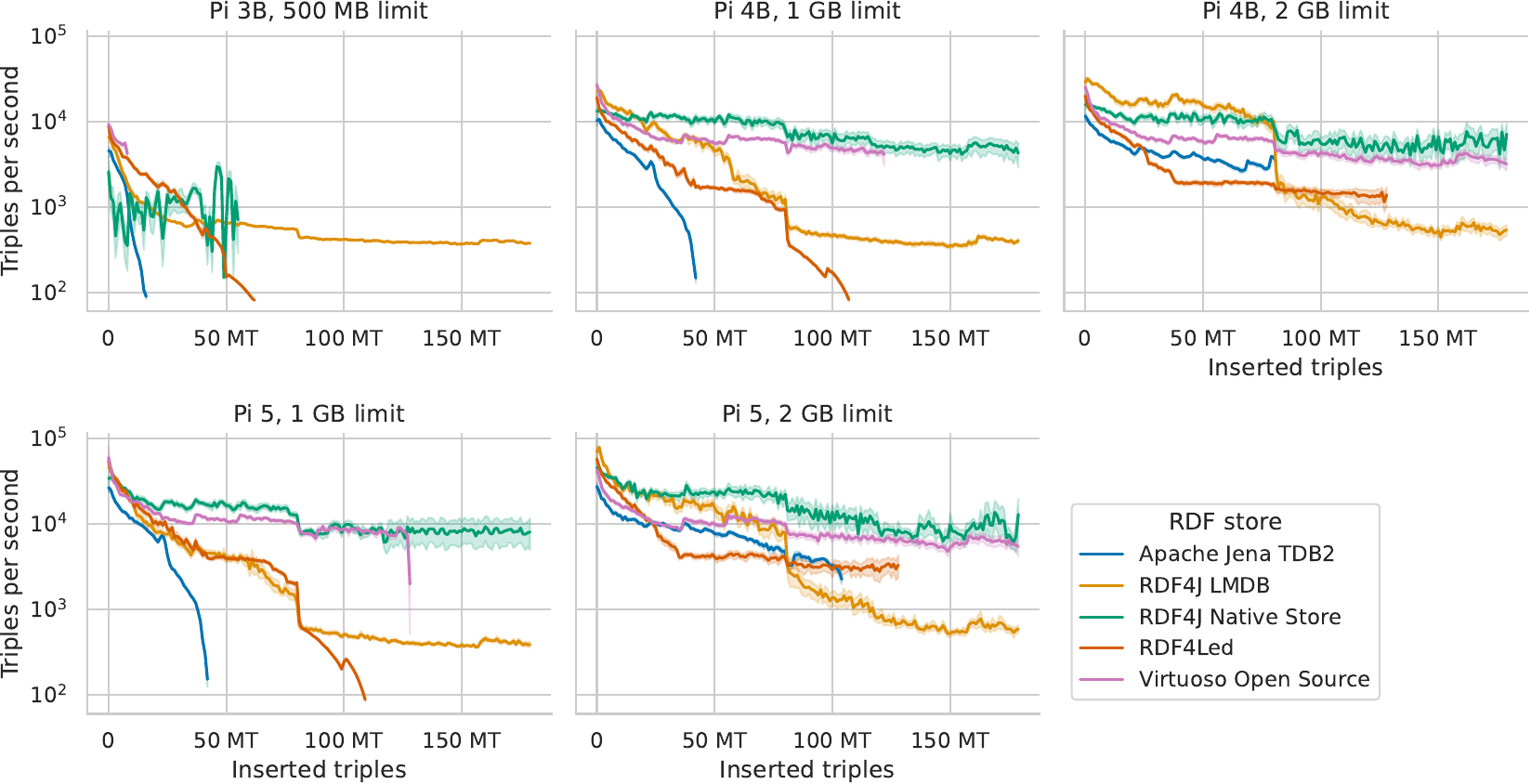}
  \caption{Loading speed comparison across different RDF stores and platforms. The results were averaged over all datasets. The Y axes are logarithmic. The data points were aggregated per 1 MT for clarity, and the shaded areas indicate the 95\% confidence interval.}
  \label{fig:store_device_speed}
\end{figure}

\vspace{-5mm}
Firstly, we compare the loading speed (in triples per second) between different RDF store implementations and hardware platforms. The results were averaged over all eight datasets, to provide general insight on typical performance of a given RDF store/platform combination. The results are presented in Figure~\ref{fig:store_device_speed}.

In the plots, lines that end prematurely indicate that the RDF store did not manage to load the dataset completely (due to a crash or the loading speed dropping below 80 triples per second). These incomplete runs are investigated in detail in the following subsection.

The first observation to be made is that Apache Jena TDB2 does not perform well in any of the scenarios, especially when the available RAM is less than 2\,GB. On the Pi 3B, until \char`\~30 million triples (MT) loaded, the fastest store is RDF4Led. This is in line with expectations, as RDF4Led was designed primarily for less powerful devices with very limited RAM. On devices with more RAM, RDF4Led crashed at \char`\~130\,MT or earlier, depending on the dataset.

All stores benefit from an increased amount of RAM. In the case of Virtuoso, this allows for loading more triples without crashing. RDF4J Native Store, on the other hand, appears to offer faster loading performance with 2 versus 1\,GB of RAM. The most interesting case is RDF4J LMDB, which entirely relies on a memory-mapped key-value store. Especially on the Pi 4B (but also on the Pi 5), moving from 1 to 2\,GB of RAM drastically improved loading performance, until \char`\~80\,MT, when the loading speed drops abruptly.

\vspace{-4mm}
\subsection{Maximum Capacity of RDF Stores}
\label{sec:results_loaded_triples}
\vspace{-2mm}

\begin{figure}[ht!]
  \centering
  \includegraphics[width=\linewidth]{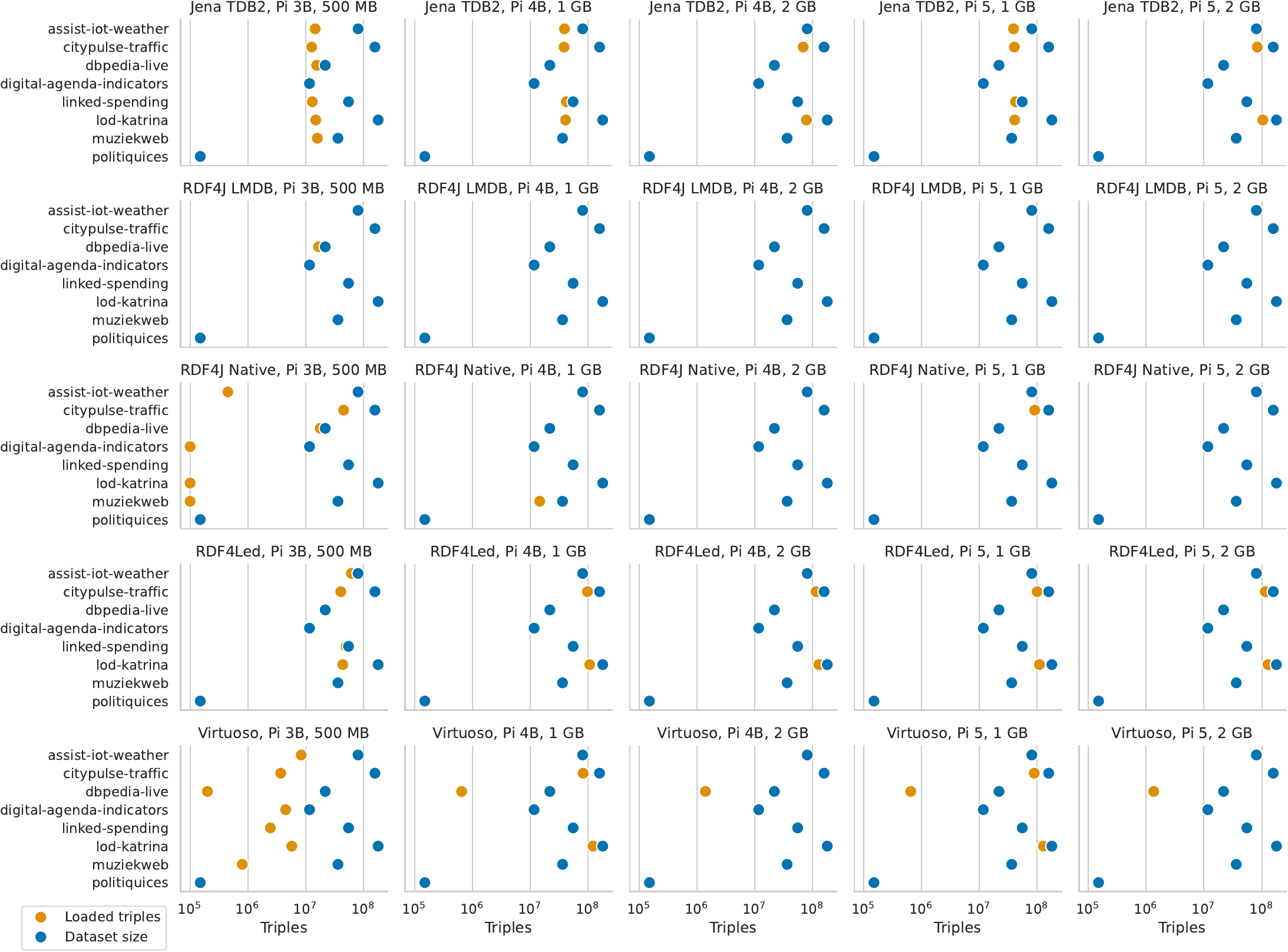}
  \caption{Comparison of dataset sizes to the number of triples that were successfully loaded by a given RDF store on a given hardware platform. If an orange dot (loaded triples) is visible, it indicates that the dataset was not loaded fully in this configuration. The X axes are in logarithmic scale.}
  \label{fig:loaded_triples}
\end{figure}

From Figure~\ref{fig:store_device_speed} it is apparent that some RDF store/platform combinations are not able to load the benchmark datasets in full. This phenomenon is presented in more detail in Figure~\ref{fig:loaded_triples}.

The most stable across all platforms is RDF4J LMDB, which only failed to fully load one dataset (dbpedia-live) on the Pi 3B. This is in contrast to the RDF4J Native Store implementation, which stopped very early on three out of eight datasets on the Pi 3B. This is due to its loading speed being very chaotic on the Pi 3B -- this is also visible in Figure~\ref{fig:store_device_speed}. For all of these three datasets, the reason for the early stop is that the loading speed dropped below 80 triples per second in one of the first batches. This is most likely due to the store not being designed to operate with so little system memory.

None of the remaining RDF stores (Apache Jena TDB2, RDF4Led, and Virtuoso) managed to load all datasets on any platform. For Virtuoso, definitely the most challenging is the dbpedia-live dataset. For TDB2, the limit appears to be strongly correlated with a specific number of triples, as on each platform all datasets stop loading around the same number of triples. A similar, but less pronounced effect can be observed for RDF4Led.

\vspace{-4mm}
\subsection{Comparison of Loading Speed Between Datasets}
\vspace{-2mm}

Finally, we investigate how does the choice of the benchmark dataset influence loading speed in general, across all RDF stores and platforms. The comparison had to first factor out the trivial influence of dataset length (in triples), which can be done by using the loading speed metric and comparing it only within a given batch of loaded triples. Secondly, one must also consider that the loading speed generally decreases as the RDF store grows in size. Therefore, the loading speed of each dataset for a given batch of triples had to be normalized to the average loading speed of all datasets for the same batch. This in turn would allow one to compare the relative loading speed between datasets. In what follows, we describe the mathematical formulas used to achieve this normalization.

Firstly, for each dataset and batch we averaged the loading speed across all available RDF stores and hardware platforms, yielding the mean loading speed (MLS):
\begin{equation}
  \mathrm{MLS}(d, b) = \frac{\sum_{s \in S} \sum_{p \in P} \mathrm{LS}(d, s, p, b)}{| (s, p) \in S \times P : \mathrm{LS}(d, s, p, b) \textrm{~is defined} |},
\end{equation}
where $d$ is the dataset in question, $b$ is the index of the batch of loaded triples, $S$ is the set of tested RDF stores, $P$ is the set of tested hardware platforms, and $\mathrm{LS}(d, s, p, b)$ is the loading speed in triples per seconds for that dataset, RDF store, platform, and batch. It should be noted here that in the calculation we omit those stores and platforms which did not manage to load this batch (LS is undefined). In the numerator of the above formula, when LS is undefined, we simply treat it as zero.

The resulting mean loading speed for a given batch was then normalized by dividing it by the mean loading speed for this batch across all datasets. This yielded the relative loading speed (RLS), which can be expressed as:
\begin{equation}
  \mathrm{RLS}(d, b) = \mathrm{MLS}(d, b) \cdot \frac{|D_{def}|}{\sum_{d' \in D_{def}} \mathrm{MLS}(d', b)}.
\end{equation}
Here, $D_{def}$ is the set of benchmark datasets for which in batch $b$ the mean loading speed is defined:
\begin{equation}
  D_{def} = \{ d \in D : \mathrm{MLS}(d, b) \textrm{~is defined} \}.
\end{equation}

\begin{figure}[htbp]
  \centering
  \includegraphics[width=\linewidth]{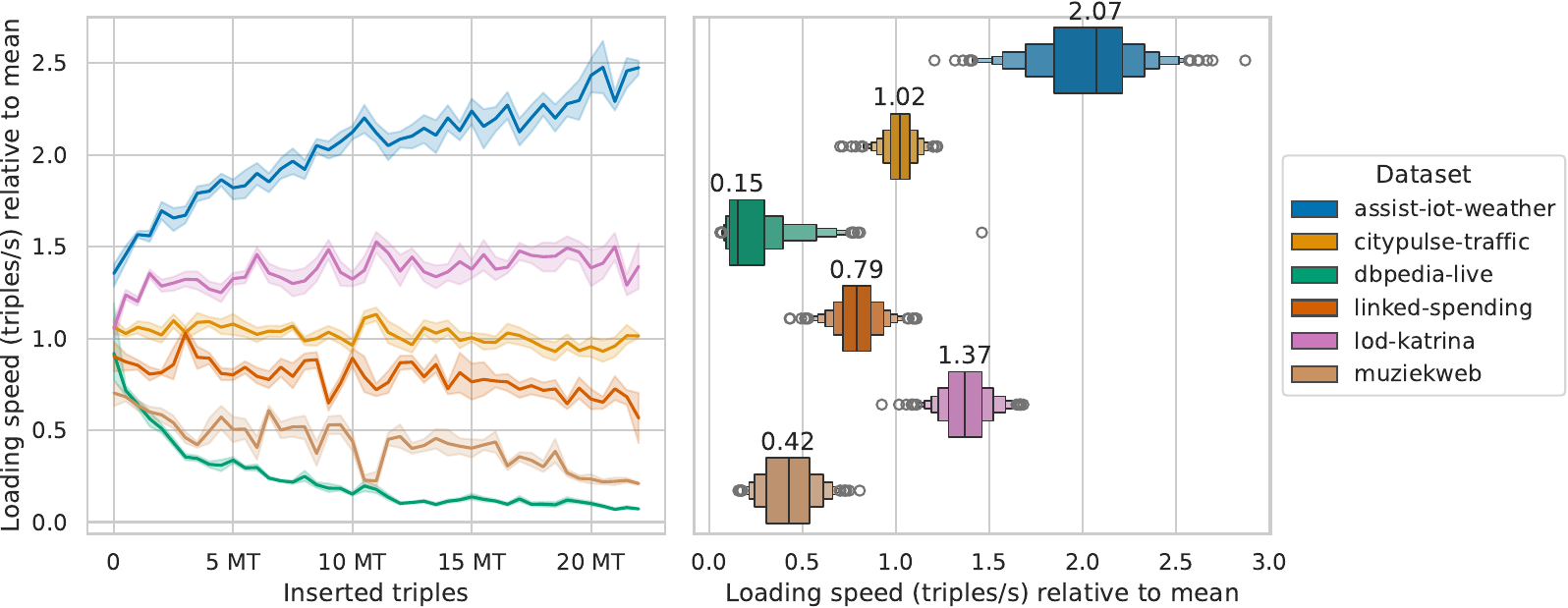}
  \caption{Comparison of relative loading speed (in triples per second) across different datasets, averaged over all device types and RDF stores. \textbf{Left:} change of relative speed over time. The data points were aggregated per 500 kT for clarity, and the shaded areas indicate the 95\% confidence interval. \textbf{Right:} distribution of relative speed for each dataset. The numerical labels indicate the median relative loading speed of a dataset.}
  \label{fig:per_dataset}
\end{figure}

With these formulas, the RLS was computed over the set of six out of eight datasets -- excluding politiquices and digital-agenda-indicators, due to their relatively small size. The remaining results were trimmed to the first 21,800,000 triples -- the length of the shortest dataset (dbpedia-live). This resulted in a set of datasets $|D| = 6$ for which in all batches the MLS was defined, so $|D_{def}| = |D| = 6$. This means that we can reliably compare the RLS across different datasets without worrying about skew due to varying dataset lengths. Figure~\ref{fig:per_dataset} presents these results.

\vspace{-5mm}
As can be seen on the left subplot in Figure~\ref{fig:per_dataset}, the relative loading speed increasingly diverges among datasets, as more triples are loaded into the RDF store. This indicates that over time, processes which depend on the size of the store (e.g., indexing) influence the loading speed more than those that do not (e.g., parsing).

The right subplot highlights that the median relative loading speeds differ drastically between datasets. The largest difference can be observed between dbpedia-live and assist-iot-weather -- the latter is loaded on average \textbf{9.01 times faster} than the former. The obtained relative speed distributions are also very regular, with few outliers and the differences between datasets clearly visible. 
By far, the slowest-loading dataset is dbpedia-live, which was also troublesome for some RDF store/platform combinations, as noted in Section~\ref{sec:results_loaded_triples}.

\vspace{-5mm}
\section{Discussion and Future Work}
\label{sec:discussion}
\vspace{-3mm}

The presented experimental results offer valuable insight into how different RDF store implementations behave on various hardware platforms with limited resources. However, perhaps the most important result is the observed wide difference in performance between benchmark datasets, across all RDF stores and platforms. After factoring out the trivial influence of the size of the dataset, we found that some datasets load faster than others by as much as a factor of 9.01. This result in turn raises the question for the reason of such drastic disparity.

In a preliminary analysis of the datasets' characteristics we found them to differ greatly in several ways: the node degree distribution, literal count, number of bytes used to encode a triple in the N-Triples format. Some of these statistics can be found in RiverBench's dataset documentation. However, none of them in isolation can fully explain the performance gap.

Although we can state with confidence that certain datasets are simply ``more difficult'' to load than others (by comparing their RLS), this level of difficulty is most likely influenced by many factors. By inspecting samples of the datasets, one can observe that, for example, assist-iot-weather (the fastest-loading dataset) has a very regular structure, with many repeating subject-predicate patterns. It may be that various cache mechanisms favor such regularity and result in very good loading performance. On the other hand, the more irregular datasets require the RDF stores to perform more changes to the indexes and cause more frequent cache misses.

It is clear that this topic should be investigated further, specifically to determine what causes a given dataset to be more or less ``difficult'' to load. We hope that the data published along with this paper can be used in subsequent statistical analyses attempting to correlate dataset characteristics with relative loading speed.

Using the already prepared and published benchmark code, the study could be extended in the future to include more RDF stores and index settings. This would allow for investigating the impact of different indexing strategies on load times. Additionally, the loading time metric could be split into the time required for parsing, and for inserting the RDF statement into the store. Although our results already show that parsing has a lesser impact on the loading time, quantifying it would yield additional useful insights and could help identify potential performance bottlenecks.

Finally, this investigation serves to show how important it is to use multiple varied datasets in RDF benchmarks. With RDF store performance varying widely between datasets, it should be considered best practice to always use multiple benchmark datasets. Additionally, the experimental results obtained here verify RiverBench's claims~\cite{sowinski2023riverbench} that its datasets are varied and thus suitable for rigorous benchmarking.

\vspace{-5mm}
\section{Conclusion}
\label{sec:conclusion}
\vspace{-3mm}

In this work, we investigated RDF loading performance across eight diverse benchmarking datasets, five RDF store implementations, and five hardware platform setups. To analyze the difference in performance across different datasets, we introduced the notion of relative loading speed (RLS). This tool is used to show that loading speed differs widely between the used datasets -- by as much as a factor of 9.01. The results clearly indicate that some datasets are more difficult to load than others, however the factors influencing this difficulty are not entirely clear. To encourage further research on this topic (especially identifying the reasons for varying loading performance across datasets), the full results of this study were published under an open license. Finally, this study highlights the importance of using multiple, diverse datasets in RDF benchmarks, and validates RiverBench as a useful and diverse resource for RDF benchmarking.

\vspace{-4mm}
\begin{acknowledgments}
  This work was partially funded by the Warsaw University of Technology within the Excellence Initiative: Research University (IDUB) program and by the Horizon Europe Research and Innovation programme under the grant agreement No. 101092908 (SmartEdge). The experiments were performed on the infrastructure of the Technical University of Berlin.
\end{acknowledgments}

\section*{Online Resources}

\begin{itemize}
    \item Processed benchmark datasets -- Zenodo, DOI: \href{https://doi.org/10.5281/zenodo.12073223}{10.5281/zenodo.12073223}~\cite{benchmark_data}
    \item Benchmark task definition -- RiverBench, PURL: \url{https://w3id.org/riverbench/v/2.0.1/tasks/flat-rdf-store-loading}
    \item Benchmark code -- \href{https://github.com/Ostrzyciel/rdf4led-riverbench}{GitHub} and Zenodo, DOI: \href{https://doi.org/10.5281/zenodo.12089254}{10.5281/zenodo.12089254}~\cite{code}
    \item Benchmark results -- Zenodo, DOI: \href{https://doi.org/10.5281/zenodo.12087112}{10.5281/zenodo.12087112}~\cite{benchmark_results}
    \item Benchmark nanopublication, PURL: \url{https://w3id.org/np/RAyFZlqsYQ_w-j5cah_gI8WBIZxiVSM4ocWHD_tnyjLxs}
\end{itemize}

\bibliography{bibliography}

\end{document}